\documentclass[useAMS]{emulateapj}
\usepackage[]{natbib}
\usepackage{graphicx}

\usepackage[colorlinks = true,
            linkcolor = blue,
                  urlcolor  = blue,
            citecolor = blue,
            anchorcolor = blue]{hyperref}

\def\lsim{\lower.5ex\hbox{$\; \buildrel < \over \sim \;$}}
\def\gsim{\lower.5ex\hbox{$\; \buildrel > \over \sim \;$}}

\shorttitle{Flaring activity in HBL 1ES 1959+650 during 2015-2016 }
\shortauthors{Kaur et al.}

\begin{document}

\title{Multi-wavelength study of flaring activity in HBL 1ES 1959+650 during 2015-16 }

\author{Navpreet Kaur$^{1, 2}$,  S. Chandra$^3$, Kiran S Baliyan$^1$, Sameer$^{1,4}$, \& S. Ganesh$^1$ }
\affil{Astronomy \& Astrophysics Division, Physical Research Laboratory, Ahmedabad, Gujarat 380009}

\altaffiltext{1}{Physical Research Laboratory, Ahmedabad, India 380009}
\altaffiltext{2}{Indian Institute of Technology, Gandhinagar, India 382355 }
\altaffiltext{3}{Tata Institute of Fundamental Research, Mumbai, India 400005}
\altaffiltext{4}{Department of Astronomy $\&$ Astrophysics, Davey Laboratory, PSU, PA 16802 USA}

\slugcomment{(Submitted to ApJ)}

\begin{abstract}
We present the results from a multiwavelength study of the flaring activity in HBL, 1ES 1959+650, during January 2015-June 2016. The source underwent significant flux enhancements showing two major outbursts (March 2015 and October 2015) in optical, UV, X-rays and gamma-rays. Normally, HBLs are not very active but 1ES 1959+650 has shown  exceptional outburst activity across the whole electromagnetic spectrum (EMS). We used the data from Fermi-LAT, Swift-XRT $\&$ UVOT and optical data from Mt. Abu InfraRed Observatory (MIRO) along with archival data from Steward Observatory to look for possible connections between emissions at different energies and the nature of variability during flaring state. During October 2015 outburst, thirteen nights of optical follow-up observations showed brightest and the faintest nightly averaged V-band magnitudes as 14.45(0.03) and 14.85(0.02), respectively. In optical, the source showed a hint of optical intra-night variability during the outburst. A significant short-term variability in optical during MJD 57344 to MJD 57365 and in gamma-rays during MJD 57360 and MJD 57365 was also noticed. Multiwavelength study suggests the flaring activity at all frequencies to be correlated in general, albeit with diverse flare durations. We estimated the strength of the magnetic field as 4.21 G using the time-lag between optical and UV bands as synchrotron cooling time scale (2.34 hrs). The upper limits on the sizes of both the emission regions, gamma-ray and optical, are estimated to be of the order of $10^{16}$cm using shortest variability time scales. The quasi-simultaneous flux enhancements in 15 GHz and VHE gamma-ray emissions indicates to a fresh injection of plasma into the jet, which interacts with a standing sub-mm core resulting in co-spatial emissions across the EMS. The complex and prolonged behavior of the second outburst in October 2015 is discussed in detail.
\end{abstract}


\section{Introduction}

Blazars are a sub-class of Active Galactic Nuclei (AGN), with a relativistic jet pointed at small angles ($<$ 15$^{\circ}$) to our line of sight \citep{urry1995}. The emission in blazars is mostly dominated by the highly variable non-thermal continuum  flux; the variability time scale  ranging from tens of minutes to a few years, across the whole  electromagnetic spectrum (EMS) \citep{bland1979, wagner1995, fan2005, fan2009}.  Their spectral energy distribution (SED) has two characteristic broad peaks implying two different emission processes at work, namely, the synchrotron process, from radio to UV/X-ray energies \citep{urrymush1982}  and the inverse-Compton (IC) process in which high energy emission (X-ray to TeV $\gamma$-rays) is produced via up-scattering of low energy seed photons by the relativistic electrons that gave rise to synchrotron emission. The  origin(s) of the seed photons are still under debate \citep{baliyan2005, bottcher2005}. According to the leptonic scenario, the IC photons may either be generated by the up-scattering of the synchrotron photons by the same population of leptons ($e^-/e^+$) \citep{konigl1981, marschergear1985, ghis-tavec2009} under Synchrotron Self Comptonization (SSC) process or the  photons from the external regions, e.g., torus, accretion disk, line emitting regions, etc., serving as seeds for the up-scattering to higher energies, under External Comptonization (EC) \citep{bottcher2007} process. On the other hand, in hadronic models, the high energy emission in blazars is mainly produced by the proton synchrotron and pion decay in the jet plasma \citep{mannheim1989}. Blazars comprise the two kinds of objects :  1.) Flat Spectrum Radio Quasars (FSRQ), identified by  emission lines in the optical/UV spectra, and 2.) BL Lac objects, identified by the extremely weak lines or a featureless optical/UV continuum \citep{stickel1993}. The classification based on the broad-band SEDs divides the BL Lac objects into three sub-categories, namely; high energy peaked BL Lac objects (HBLs;  $\nu_{s} \textgreater 10^{15} $ Hz ), Intermediate energy peaked BL Lac objects (IBLs;  $ 10^{14} Hz \textless \nu_{s} \textless 10^{15}  $ Hz) and Low energy peaked BL Lac objects (LBLs;  $ \nu_{s} \textless 10^{14} $ Hz). The high energy emission in the HBLs is generally well explained by the SSC models, with a possible EC component in a few exceptional flaring states \citep{bottcher2007}. 
     
    The blazars being extremely variable across the EMS, their variability can serve as a tool to understand AGN structure and emission processes, as their central engines are too compact to be  resolvable \citep{ciprini2003, marscher2008}. The HBLs generally are less variable than LBLs \citep{jannuzi1994} but some of them are very active with flares and outbursts detected almost over the complete accessible EM spectrum, ranging from the radio to TeV $\gamma$-rays \citep{acciari2011, furniss2015}. 
 
While almost all the TeV flares are witnessed to have a counterpart in optical and X-rays, barring some orphan flares,  the GeV energy region might show weaker activity. Blazars also show significant polarization in optical \citep[ and references there-in]{chandra2011} and radio wavelengths which is a measure of the alignment and the strength  of the magnetic field. The changes in the degree of optical polarization (DP) and position angle (PA) are commonly seen during the flares in the blazars. Such rapid variations in  DP and PA during a flare have been modeled for many sources (e.g., for 1ES 1011+496 - \citet{aleksic2016}, Mrk 421 - \citet{zhang2015}, for 3C 279 - \citet{kiehlmann2016, hayashida2012, abdoPol2010}). The outbursts in  blazars are mostly thought to be a manifestation of the shock formation and their movement down the jet \citep{orienti2015, marscher2010},  internal inhomogeneities and their interaction with shocks,  re-collimation of shocks downstream the jet causing re-acceleration \citep{spada2001} or, a new population of the relativistic plasma injected into the jet. In spite of the considerable efforts until now, none of the proposed models are able to explain blazar phenomena. Our understanding of the geometry of the jet, the emission processes responsible for different flaring activities and the behavior of the objects during their quiescent phase are limited by the sample size and the scarcity of simultaneous data over a broad energy range. Therefore, there is  need for extensive multi-wavelength studies on a large sample of blazars to enable a comprehensive understanding of the emission processes, in general.

The HBL 1ES 1959+650, redshift z=0.048 \citep{perlman1996}, was first detected in radio band using NRAO Green Bank Telescope \citep{gregorycondon1991} and observed in X-rays during the slew survey by the Einstein Imaging Proportional Counter \citep{elvis1992}. The first TeV detection of this source was reported by the Seven Telescope Array group in 1999 \citep{nishiyama1991}. This source was identified as an optical BL Lac object by \citet{schachter1993}. Later, a  bright ($m_{R}$ = 14.9) elliptical galaxy was confirmed by \citet{scarpa2000} as the  host. The source has undergone various outburst stages, including intense activity at very high energies (GeV-TeV). \citet{krawczynski2004} reported an "orphan" flare at VHE during an outburst in 2002, in a multi-wavelength campaign  (WHIPPLE and HEGRA for TeV, RXTE for X-rays, Boltwood and Abastumani observatory for optical, UMRAO for radio 14.5 GHz) from May 18 - August 14, 2002.  The authors reported a correlation between the $\gamma$-ray and X-ray fluxes but during orphan TeV flare, no enhancement in X-ray flux was seen and  there was no correlation between  optical and X-ray/$\gamma$-ray emissions. \citet{bottcher2005} explained 2002 orphan TeV flare using a hadronic synchrotron mirror model in which the orphan TeV photons originated from the interaction of relativistic protons with an external photon field supplied by synchrotron radiation reflected off a dilute reflector.  Another intense flaring activity was seen during 2012 April- June covered in a multi wavelength campaign  by \citet{aliu2014}. During the outburst, 1ES 1959+650 emitted enhanced flux in gamma-rays without any significant simultaneous rise at X-ray energies. The authors proposed a reflected emission model to explain elevated $\gamma$-ray flux, via pion production with very high energy protons (10-100 TeV). 
 
	1ES 1959+650 was reported in unprecedented high flux state across all the energies (gamma-ray, X-ray, UV, Optical and radio) during 2015 which extended to 2016 as well. The source  underwent two major outbursts; first in X-rays during March 2015 and the second one during October 2015 with  exceptionally large count rates (more than 20 counts/sec) in X-rays \citep{AtelXray2015}, making it only  the third TeV source with such high X-ray count rate after Mrk 421 and Mrk 501. Such outburst activities provide an opportunity to study the underlying physical processes responsible for emission at different energies. Recently, \citet{kapanadze2016} reported prolonged X-ray activity in 1ES1959+650 during their 6-month coverage of 2015 October outburst (MJD 57235-57410) along with enhanced $\gamma$-ray (0.3-100 GeV) flux at several epochs. They also claimed an orphan $\gamma-$ray flare at MJD 57314 with no enhancement  at other energy regimes, albeit with a rider that due to sparse data, a rapid X-ray flare might have occurred.   To understand the overall behavior of 1ES 1959+650, we carried out an extensive multi-wavelength study of the outburst activities in 1ES 1959+650 from 2015 January-2016 June (MJD 57040-57570). Optical follow-up observations from Mt Abu InfraRed Observatory (MIRO) were carried out for the source when it exhibited high flux state in X-ray, GeV and TeV energies. This paper is organized as follows. Section 2 describes the observations and data analysis techniques in optical, X-ray and Gamma-rays. Detailed analysis of the light curves and the results are discussed in Section 3. Section 4 presents a brief summary of the work presented in this paper.

\section{Multi-wavelength Observations and Data Reduction
 \label{sec:sec2}}

Multi-wavelength data from various resources, space-borne observatories, namely, Fermi ($\gamma$-rays), Swift (UV, optical  \& X-rays) and the ground-based facility, MIRO (optical/IR) are used in this study. We have also used  publicly available archival data from Steward Observatory, Arizona \citep {smith2009}in optical  and radio data at 15 GHz  from OVRAO \citep{richard2011}  to discuss various outburst episodes. We briefly summarize the data analysis techniques used for the data from aforementioned resources in the following. 

\subsection{Gamma-ray : Fermi-LAT }

FERMI Large Area Telescope (Fermi-LAT) is a primary instrument on board the Fermi satellite \citep{atwood2009}. The LAT has an unprecedented sensitivity in the $\gamma$-ray band (20 MeV - 300 GeV) and scans the entire sky in approximately  3 hrs except for few extremely high priority specific pointing mode observations where the observations are taken for 30 minutes for the prioritized sources. It provides a multi-dimensional data base of location, energy and time for each detected event. 

We analyzed  1ES 1959+650 Fermi-LAT data from January 01, 2015 (MJD 57023) to June 31, 2016 (MJD 57550) using standard recommended procedure by making use of the latest ScienceTools (version v10r0p5). The photon class events lying within the region of interest (ROI) of 10$^{\circ}$ , zenith angle $\textless$ 100$^{\circ}$ , within the energy range of 0.1 - 300 GeV are extracted using "gtselect" tool. We discarded the data when the rocking angle of the spacecraft was greater than 52$^{\circ}$ to avoid any photon contamination from the Earth's limb. An unbinned likelihood analysis was performed using gtlike tool with the help of input source model covering a region of  20$^{\circ}$ around the source position, generated using 3rd FGL catalog \citep{acero2015}.

A maximum likelihood analysis using gtlike has been used to reconstruct the source energy spectrum. The background model was constructed using third Fermi LAT catalog $(gll_psc_v16.fit)$ that contains 36 gamma-ray sources lying within ROI \textless 12$^{\circ}$, as well as diffuse emission with no extended sources within this region. We have made use of a log-parabolic model and a power law model for the sources with significant and without spectral curvature, respectively. The source (1ES 1959+650) spectral parameters within 3$^{\circ}$  were kept free during spectral fitting, while sources outside of the aforementioned range were held fixed as in the 3FGL catalog. The Galactic diffuse emission and the isotropic emission component was modeled using ${\textit gll\_iem\_v06.fits}$ and $iso\_P8R2\_SOURCE\_v6\_v06.txt$, respectively. Fermi-LAT data was reduced using a Python based package called Enrico \citep{sanchezdiel2013}. A time binning of 2 days was used to extract the source light curve. To look for the details of post-outburst activity in the source, we used 2.5 day binned data set.

\subsection{X-ray, UV, optical : Swift-XRT/UVOT}

We have made use of around 95 observation IDs observed by the instruments XRT and UVOT onboard Swift,  during January 01, 2015 to December 31, 2015. The {\it heasoft} (version 6.17) package along-with the recently updated calibration database (2016 January 21 for XRT \& 2016 March 05 for UVOT) is used for the analysis of the above-mentioned data.

The {\it xrtpipeline} tool provided freely as a part of {\it heasoft} package, with default parameters are used to extract the cleaned events files. This source, being very bright, is mostly observed in WT mode. The typical full frame count rates for WT mode observations are always less than 25 c/s which will be pile-up free as recommended by the instrument  team at University of Leicester, UK  \footnote{http://www.swift.ac.uk/analysis/xrt/xselect.php} (pile-up occurs for rate $\ge$ 100). Following this, the pile up corrections are not performed for WT mode data. For this case, a circular area of 27 pixels (equivalently $\sim$ 63") centered at the position of the 1ES 1959+650 is used as the source region. The background region was extracted as a concentric annulus with inner and outer radii of 80 pixels and 120 pixels, keeping the average half width of annular region at 100 pixel (recommended for a proper background subtraction in WT mode). The PC mode observations are corrected for pile up using the prescriptions suggested by aforementioned team \footnote{http://www.swift.ac.uk/analysis/xrt/pileup.php}. The innermost circular area with radius 10" is excluded (chosen for a highly piled up observations) from the source region of 70" circle around the source. The background in this case is used as an annular region around the source, with inner and outer radii of 150" and 350", respectively.  The clean events files are then used to extract the products (spectrum \& light-curves) for the source and background regions using the  {\it xselect} tool. 

\begin{figure*}
\centering
\caption{Multi-wavelength light curve (MWLC) for 1ES 1959+650 from January 2015 to June 2016, showing three major flaring episodes. Flare 1 (March 2015),  Flare 2 (October 2015) and Flare 3 (June 2016) in $\gamma-$rays are clearly visible in the MWLC, where time in MJD is plotted along X-axis and Y-axis is the fluxes/magnitudes. Top panel shows Fermi-LAT fluxes with 3$\sigma$ statistical error bars. The second panel is soft X-ray (XRT-Swift) with three energy bands (0.3-3.0 KeV, 3.0-10 KeV, and 0.3-10 KeV). Third panel shows UV flux (in mJy; W1 band) from UVOT onboard Swift. Fourth  panel display optical V-band magnitudes obtained from Swift, MIRO and Steward Observatory. Last panel shows radio flux at 15 GHz taken from OVRO public archive. }
 \label{fig:fig1}
\includegraphics[width=18cm,height=14cm]{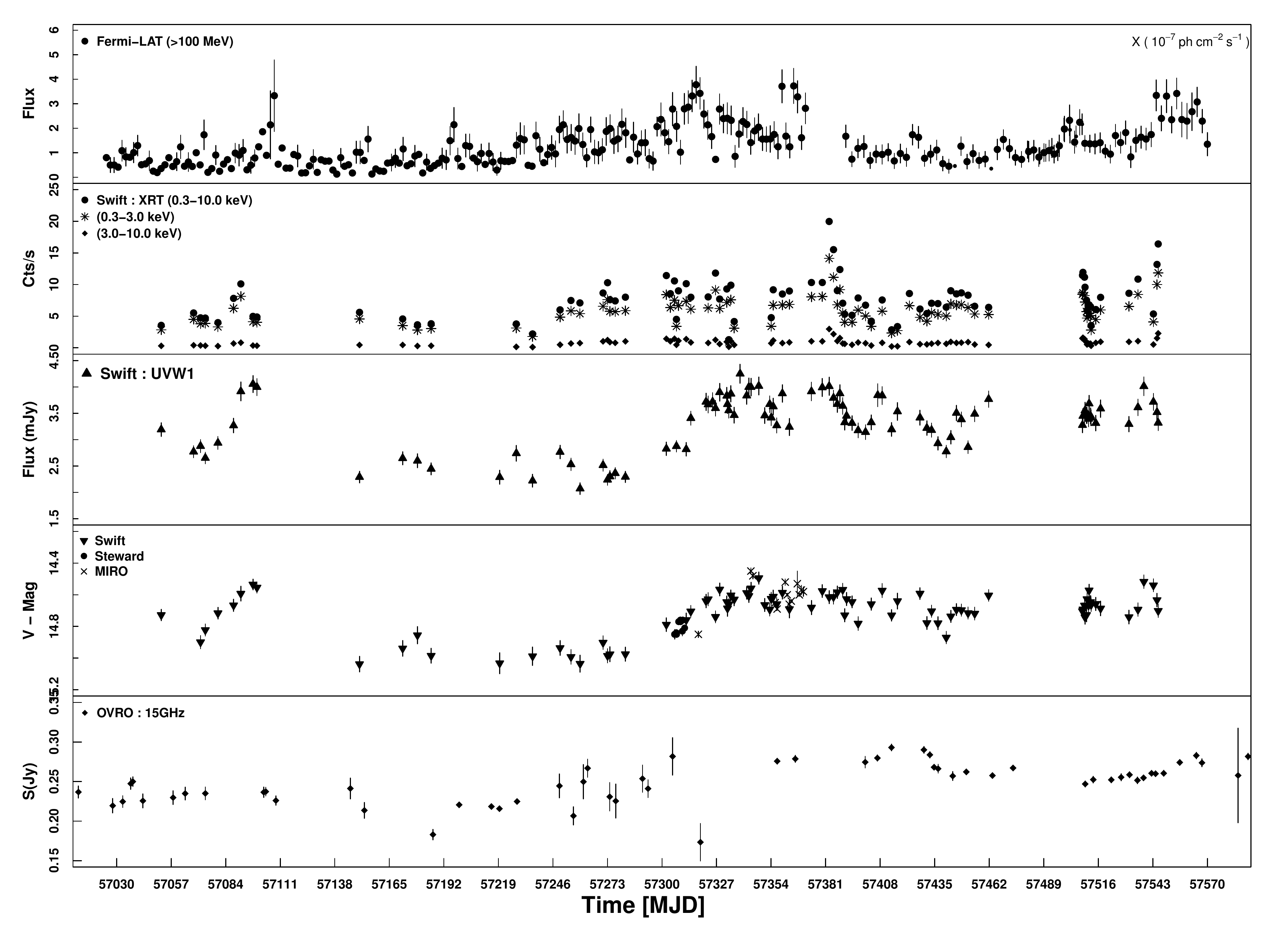}
\end{figure*}

The spectra thus obtained are then fitted with an absorbed log-parabola model with nH value fixed to the galactic value (1.07 $\times$ 10$^{21}$ cm$^{-2}$) using {\it xspec} (version 12.9.0), a standard tool for X-ray spectral fitting provided as a part of {\it heasoft} package.  
Using the BACKSCAL keyword in WT mode, source and background spectrum files were edited to the proper values, before importing to the fitting tool, to avoid the wrong background subtraction during the fit. The log-parabola was chosen instead of the commonly used absorbed power-law which was giving very poor fit ($\chi_{\nu}$ $\ge$ 1.9). The absorbed broken power-law model was providing similar fit as that using curvature model. We prefer log-parabola because it provides  natural turn-over in the spectrum instead of a sudden break as given by broken power-law models. The background subtracted count rates extracted for energy band 0.3-10.0 keV are used to generate the light-curves (see Fig. \ref{fig:fig1}). The unabsorbed fluxes in 0.3-10.0 keV band are also estimated by adding component ``cflux" and fitting after freezing the normalization. The fluxes thus obtained are also used for timing analysis in this paper. 

The UVOT data analysis is done in a similar fashion as adopted in \citet{chandra2015}. The snapshots observations in the filters V (5468 \AA), B (4392 \AA), U (3465 \AA), UVW1 (2600 \AA), UVM2 (2246 \AA), and UVW2 (1928 \AA), for all the OBsIDs, were integrated with the {\it uvotimsum} task and analyzed using the {\it uvotsource} task, with a source region of 5'', while the background was extracted from an annular region centered on 1ES 1959+650 with external and internal radii of 40'' and 7'', respectively. The observed magnitudes from all OBsID are then corrected for extinction according to the model described in \citet{cardelli1989}. A tool, developed in-house, using R-platform\footnote{https://www.r-project.org/}, is used to perform the required reddening corrections. The corrected fluxes are then used for timing analysis in the  following section \ref{sec:sec3} (See Fig. \ref{fig:fig1} for light curve).

\subsection{Optical observations : MIRO}

Following an alert \citep{Atel8193} of an enhanced $\gamma$-ray  activity in 1ES 1959+650 on October 20, 2015 (MJD 57315), we made optical photometric observations using two telescope facilities at Mt. Abu InfraRed Observatory (MIRO) i.e., 1.2 m \&  0.5 m telescopes. The 1.2 m telescope is equipped with  LN2-cooled CCD (1296 $\times$ 1152 pixels; pixel size = 22 $\micron$) at its f/13.2 Cassegrain focus, whereas, a thermo-electrically cooled (T $\sim$-80$^{\circ}$C)  iKon ANDOR CCD (2048 $\times$ 2048; pixel size =25 $\micron$) is used as backend instrument for 0.5 m telescope. The dark current in both systems is negligible. 

 The observations were carried out using  BVRI Johnsons-Cousins filters for total thirteen nights during October 23 to December 13, 2015. High temporal resolution (exposure time of $\approx$ 40-50 seconds) data with  the optimum signal-to-noise ratio (S/N $\textgreater$ 5) are obtained. The flat field images were taken during twilight and dawn on daily basis whereas bias frames were taken whenever telescopes were slewing to change the source. During 2015 November- December, we also monitored the source to look for intra-night variability (INV).   
   
The data were reduced using standard data reduction procedures using IRAF package (Image Reduction and Analysis Facility) and locally developed pipelines  \citep{chandra2011,Nav2017}. On each night, master bias and master flat frames were generated by combining all bias and flat field images, respectively. The science images were then corrected with bias and flat field and aperture photometry was performed on the source as well as on comparison stars using DAOPHOT package. The source magnitudes thus obtained were calibrated using two comparison stars 4 and 6 \citep{villata1998} having similar brightness as that of the source. 

 Table \ref{tab:table1} gives the details of photometry data obtained from MIRO where column 1 and 2 represent the  date and MJD of observations, third and fourth column are nightly averaged R- and V-band magnitudes along with photometric errors, column 5 shows the telescopes used. In order to look for INV  in 1ES 1959+650, we monitored the source for more than  an hour on 9 nights during October - December 2015.

\subsection{Supplementary data : Optical (Steward Observatory);  Radio(OVRO)}
We have used optical photometry and polarimetry data available online from Steward Observatory \footnote{http://james.as.arizona.edu/~psmith/Fermi/} during October 2015 to look for the polarization behavior during its flaring state. We have also utilized the publicly available radio data from Owens Valley Radio Observatory (OVRO)\footnote{http://www.astro.caltech.edu/ovroblazars/index.php?page=home} at 15 GHz frequency  to checkfor  any correlated activity with other data used here.

\section{Results and Discussion}
 \label{sec:sec3}
Multi-wavelength light curves (MWLC) are constructed using analyzed data as described in the previous section  (section 2) and are shown in Figure \ref{fig:fig1} where X-axis represents the time in MJD (Modified Julian Day) and Y-axis shows respective  flux or magnitude values at various energies.
In Figure \ref{fig:fig1} (from the top), the first panel: Fermi-LAT $\gamma$-ray (0.1 to 300 GeV) flux, second panel:  Swift-XRT flux  at three X-ray energy bands i.e., (0.3 - 3.0 keV: X1 band), (3.0 - 10.0 keV : X2 band), and (0.3 - 10.0 keV : X3 band), third panel: Swift-UVOT UV (UVW1-band) light-curve, fourth panel: Swift-UVOT, MIRO  and Steward Observatory V-band optical light-curve, fifth panel: OVRO 15 GHz radio light-curve.  
		
\subsection{Multi-wavelength light curve}
\label{sec:sec3.1}
As can be noticed from the Figure \ref{fig:fig1},  the light curves across all the energy bands (Gamma-rays -- X-rays -- UV -- Optical -- Radio) appear very complex in nature, especially during the two major outbursts, with a number of flares, sub-flares with diverse rates and periods of quiescence appearing through out the  entire electromagnetic spectrum (EMS). Such random trends are typical in blazar light-curves \citep{Chatterjee2012}. The shape and length of flares in the light-curve along with polarization information tell about the emission mechanisms at work, strength of the magnetic fields, etc., in the jet. A rising trend in MWLC corresponds to the acceleration of the relativistic particles as a dominant process while a declining trend indicates to their subsequent cooling.
  
   The flux at GeV energies (0.1 - 300 GeV) for 1ES 1959+650 remained in low state (F$_\gamma$ $\approx$ 1.3 $\times$ 10$^{-08}$ ph\  cm$^{-2}$ s$^{-1}$) most of the time super-imposed by several mini-flares during 2015-16. It should, however, be noted  that it is more than twice  the average flux level reported in 3FGL catalogue (F$_\gamma$ $=$ 5.83 $\pm$ 0.18 $\times 10^{-09} ph\ cm^{-2} s^{-1}$). During the October 2015 outburst, the source showed  flux level as high as  3.8 $\times$ 10$^{-07}$ ph cm$^{-2}$ s$^{-1}$ with 2-day binning, highest ever reported for this source in the (0.1-300 GeV) range. \citet{kapanadze2016} (hereafter, K16) reported $\gamma$-ray flux  as 1.2$\times$ 10$^{-07}$ ph cm$^{-2}$ s$^{-1}$ at this epoch  in the (0.1-100) GeV range with a 3 day binning. The difference in maximum flux level  might have arisen due to different binning, energy range covered (0.3-100 GeV) and analysis method used in K16.
   
   Several sub-flares, with varying rise/fall rates, have been noticed before and after the onset of major outbursts - March 2015 (outburst 1) and October 2015 (outburst 2) which is extended to 2016 June. On the other hand, X-ray flux was seen to behave erratically during the whole period with significantly enhanced flux when the source was in outburst in $\gamma-ray$, UV, optical and radio. A flare beginning at MJD 57070 (outburst 1) appears truncated due to lack of data in X-ray; UV and optical met the similar fate while a clear flare is noticed in $\gamma$-ray light-curve, peaking at about MJD 57107 (F $\approx$ 3.2 $\times$ 10$^{-07}$ ph cm$^{-2}$ s$^{-1}$).  The trend shows that a complete data set could have led to correlated flaring in all these energy bands.

  In soft X-ray band (0.3-3.0 keV; X1 band) counts varied from 3-13 cts sec$^{-1}$, but for X2-band (3.0-10.0 keV; relatively hard flux state), the counts remained between 1-2 cts sec$^{-1}$. During the outburst phase, the total number of counts in X-ray reached as high as 10 - 20 cts sec$^{-1}$. However, K16 report maximum count rate as 22.95 in 0.3 - 2.0 keV range at MJD 57382.8, same as reported in \citet{K16_ATel2015}. Such high X-ray count rate makes it only the third TeV source after Mrk 501 \& Mrk 421. The UV light curve (3rd panel  from the top) showed significant variation during outburst period. A consistently high flux is seen starting around MJD 57245 (2015 August 10: onset of outburst 2), with signatures of various flares spanning a few days, super-imposed over an already high flux. Almost similar trend is seen in optical V-band light curve but with less flux modulations superimposed on the increased flux level. These elevated flux levels, almost twice the base value, continued for more than 200 days due to flaring activity. The radio light curve at 15 GHz shows slowly decreasing flux up to MJD 57184 after which it starts increasing again, showing strong flare and then crossing 2.5 Jy level at about MJD 57304 (2015 October 9: outburst 2).  After this, radio flux drops sharply, reaching lowest value in the whole duration, just when optical degree of polarization and position angle had undergone rapid changes (cf Figure \ref{fig:fig2}, right panel). Though there is very limited data on DP and PA, these  rapid variations, are significant as they are closely followed by a major flare in gamma-rays. Subsequently,  optical, UV and X-ray emissions show their peak in their respective light-curves. The details of the second outburst will be discussed in the context of $\gamma$-rays, in the next section.

Apart from these major flaring activities, several mini-flares spanning a few days in $\gamma$-rays can be seen in the light-curve, e.g., flare peaking at about MJD 57070 (flux drops from (1.7- 0.2) $\times10^{-7} ph cm^{-2}s^1$), 57152 (with simultaneously enhanced emissions at radio, UV \& X-rays), 57191 (no data in X-ray/UV/Optical but 15GHz flux is significantly enhanced), 57340 etc. The detailed investigation of the light-curve reveals twin peaks in $\gamma$ $-$ rays, spanning almost 6 days each, just after the major outburst 2 peak. Both the flares are at almost the same flux level as that of the outburst 2.

In the following section, we will discuss various features in the light-curve.

\subsection{MWLC: Outburst 1 (March 2015)}

The duration of March 2015 outburst  in  $\gamma$-ray flux was roughly of 29 days (MJD 57094 - 57123) with a peak flux value reaching  $F_{\gamma} $ = 3.25$\times$ 10$^{-07}$ ph cm$^{-2}$ s$^{-1}$ on MJD 57109. During the peak outburst, source brightened by more than 10 times the quiescent state flux value  ($F_{\gamma} \textgreater$ 0.3 $\times$ 10$^{-07}$ ph cm$^{-2}$ s$^{-1}$). The outburst is temporally almost symmetric with rise time and decay time as 15 and 14 days, respectively.

The outburst has several pre-burst flares contributing to it with  diverse rates. Similarly, outburst in X-ray has a duration of about 17 days, peaking around MJD 57091 with 10.1 counts $s^{-1}$.  The rising rate for X-rays is 4.4 $\pm$ 1.61 $\times$ 10$^{-12}$ ergs cm$^{-2}$ s$^{-1}$ with a rather sharp decline. However, it is quite possible that it could have been truncated due to lack of the data, the peak still to occur, just like what we see in UV and optical bands. In that case, all the emissions are likely to peak simultaneously, indicating to having almost same origin.

 Notice the similar enhancement in the form of a mild outburst in radio band  lasting for about  14 days with a sharp decay. It appears that March 2015 outburst activity first started in radio band, then in X-rays, UV, optical and later in $\gamma$-ray. However, there are some pre-outburst flares seen in optical, UV and gamma-rays as well. The $\gamma$-ray outburst starts when X-ray emission has almost peaked and UV/optical are in the process of reaching peak values (Figure \ref{fig:fig1}). X-ray peak leads UV/optical and $\gamma$-ray ones by about 7 and 8 days, respectively. Since 1ES 1959+650 is an HBL, X-ray emission (dominated by soft X-rays) is expected to come from  synchrotron process which is also responsible for UV and optical emissions.  High energy particles giving rise to X-ray emission cool faster and hence X-ray emission leads optical/UV. However, 7-days lag between X-ray and optical/UV appears to be a bit long and perhaps emission regions are also aligned differently to LOS. It could also be that UV and optical peaks could not be seen due to truncated data.  In SSC, the $\gamma-ray$ emission is produced by IC process where synchrotron seed photons take time to travel to high-energy ($\gamma-ray$) emission region, where these are up-scattered \citep{sokolovMarscher2004}.  It is possible that the emissions are generated when the emitting plasma passed through a standing or slowly moving shock down the jet and below the sub-mm core.  \citet{jorstad2001} reported that $\gamma$-ray flares were associated with ejection of super-luminal radio knots into the jet, which initiates flare in radio and $\gamma$-ray flare followed.

 All the four flares in the light-curve upto MJD 57220 rise slowly while decay very fast. The variations during the outbursts clearly follow a common trend across all the energies, indicating that acceleration timescales for electrons are longer than their cooling timescales. 

\subsection{MWLC: Outburst 2 (October 2015- June 2016)} 

The build up to the second major outburst (MJD 57285 - 57370 or upto 57570) started much earlier, due to lack of data in other bands (X-ray, UV and optical) we consider the period from MJD 57225,  which marks significant enhancement in the fluxes in all energy bands, including 15GHz radio band as shown in Figure \ref{fig:fig2}. While flux in optical, UV and $\gamma-$ray starts gradually, flux in radio and X-ray energy bands rise sharply, reaching a plateau at about MJD 57246 where even $\gamma-ray$ flux has also increased significantly, while optical and UV are still rising. These two bands reach plateau when radio and $\gamma-$ray fluxes have already peaked.

 The rise and fall rates for $\gamma-$rays during 2nd outburst were estimated as - 9.01 $\pm$ 2.21 $\times$ 10$^{-09}$  ergs cm$^{-2}$ s$^{-1}$ (MJD 57285 - 57317) and -3.31 $\pm$ 1.08 $\times$ 10$^{-09}$ ergs cm$^{-2}$ s$^{-1}$ (MJD 57317 - 57326). For X-rays the estimated  rise rate is 3.44 $\pm$0.82 $\times$ 10$^{-12}$  ergs cm$^{-2}$ s$^{-1}$ (MJD 57249 - 57382). The main outburst showed sudden flux enhancement, supported by sub-flares, spanning 41 days in $\gamma$-rays  (MJD 57285 - 57326) reaching its peak value on MJD 57316.8 (2015 October 22). A prolonged, erratic flaring activity (with more than twice the average flux) with duration of about 145 days in X-ray flux is clearly seen, in which a flare peaks around MJD 57382 with largest ever 20 cts s$^{-1}$, followed by four other significant flares, each of them detected with \textgreater 10 cts s$^{-1}$. Note that we have selected X-ray data-sets to estimate flare duration when the source count rate was above 5 cts s$^{-1}$.
 
Let us discuss the major flares in the light-curve where $\gamma-ray$ flux is normally above $ 10^{-7} ph cm^{-2}s^{-1}$. The flare at MJD 57237 decays by 1$\times 10^{-7} ph cm^{-2}s^{-1}$ in 2d, rising again by about 1.2$\times 10^{-7} ph cm^{-2}s^{-1}$ to 1.7$\times 10^{-7} ph cm^{-2}s^{-1}$. The flux in two other flares at MJD 57247.2 and 57262 rise by more than two-fold to about 2.0$\times 10^{-7} ph cm^{-2}s^{-1}$ in 2d. The pre-outburst flux of flare at MJD 57297 changes from  0.7$\times 10^{-7} ph cm^{-2}s^{-1}$ to 2.14$\times 10^{-7} ph cm^{-2}s^{-1}$ and rises further to 2.4$\times 10^{-7} ph cm^{-2}s^{-1}$ at MJD 57302. It is followed by peak flux (0.28Jy) in radio at MJD 57305 and a 26$\sigma$ detection of VHE by VERITAS \citep{mukherjee2015_atel_8148}, accompanied by a significant enhancement in optical, UV and X-ray flux. After two more flares contributing to the flux, October 2015 outburst reaches  its peak flux 3.75$\times 10^{-7} ph cm^{-2}s^{-1}$ at MJD 57316.88 (2015 October 22) in the (0.1-300 GeV) energy range. The peak in $\gamma-$ray flux is followed by delayed peaks in optical, UV, and X-ray bands showing significant variability at about MJD 57322. However, radio flux is at its lowest after a sharp decay. It should be noticed that this major flare in $\gamma-$rays happened just after rapid changes in optical polarization and position angle followed by peak in radio flux. All these activities signal that the emissions were correlated and were perhaps caused by passage of a blob through the sub-mm core as around the same time significant VHE emission and noticeable radio emission \citep{mukherjee2015_atel_8148, RATAN_Atel2015}  was detected . We, therefore, feel that the  occurrence  of an orphan flare in the (0.3-100 Gev)  $\gamma-$ray range is doubtful at this epoch, as claimed by \citet{kapanadze2016}, while discussing the X-ray flux variations with other bands.

After the major flare, the $\gamma-$ray flux drops to 0.7$\times 10^{-7} ph cm^{-2}s^{-1}$ in 10d, rising again to 2.7$\times 10^{-7} ph cm^{-2}s^{-1}$ in 2d. This flare at MJD 57328 is followed by X-ray (K16), UV and optical. The $\gamma-$ray flare at MJD 57342 is accompanied by E1-event in X-ray as reported in K16 (we do not have that data as it was a ToO observation) with 20 cts/second, followed by first of twin peak flare in optical, discussed later and flare in UV. FACT also reported 3$\sigma$ detection of VHE emission at this epoch. The first twin $\gamma-$ray flare at MJD 57360, flux 3.5$\times 10^{-7} ph cm^{-2}s^{-1}$, is followed by, after one day, second optical twin peak and preceded by E3 event in X-ray (19 cts/s; K16). Radio emission is also enhanced with enhancement in UV flux. All the emissions appear to be nicely correlated. After this, there is break in our $\gamma-$ray data, but there are significant flares in optical, UV and X-ray. In fact this was also ToO slot as reported in K16, where a $\gamma-$ray flare is accompanied with E2 event in X-ray along with enhanced flux in optical and UV. 

A nicely correlated flare in radio, optical, UV, X-ray and $\gamma-$ray fluxes is noticed  peaking at MJD 57421, in the light-curve followed by a Swift data break upto MJD 57510. Just before that, a large outburst occurs (flux change (1 - 2.2) $\times 10^{-7} ph cm^{-2}s^{-1}$) at MJD 57506, decaying part of which is captured by all other bands.
A clean flare in UV, optical, followed by X-ray and $\gamma-$ray flares at MJD 57540, with 0.25 mag brightness in V and a change in UV flux 2.5-3.5 mJy is seen. A number of flares contribute to the  last outburst in $\gamma-$ray flux centered at MJD 57544.72 when flux increases almost three fold to 3.34 $\times 10^{-7} ph cm^{-2}s^{-1}$ , the outburst lasting about  25 days. X-ray counts increase from 5 to 15, radio  also shows correlated enhancement in the flux while UV and optical are decreasing.
  
  As evident from the 15GHz radio light-curve, the source was active much before the enhancement started in X-ray/UV/optical emissions, i.e., around MJD 57185 and exhibited slow rise, peaking around MJD 57305 with a flux value of 0.28 $\pm$ 0.02 Jy, and  a sharp fall reaching to half of its peak flux in 14 days, i.e., 0.17 $\pm$ 0.02 Jy on MJD 57319, extinguishing the 134 days of activity in radio. This sudden "shut off'' of activity in radio happened just 7-days prior to the highest peak in $\gamma$-ray flux and just after the rapid changes in the optical polarization (see, Figure \ref{fig:fig2}) had taken place. \citet{Lahteenmaki2003, jorstad2001} also noticed flaring of radio emission at 37 GHz just before a flare in  $\gamma$-ray flux occurred. With a break in the data, enhanced flux, by a factor of two ($F_{radio} = $ 0.250 $\pm$ 0.002 Jy), is noticed in 15GHz radio band.  

 Therefore, across the whole spectrum, an outburst with significant high flux levels and  slow rising/decaying trend spanning over a few months (long-term variations) is noticed along with flux variations lasting for a few days (short-term variations). A rapid variability in $\gamma$-rays with clear mini-flares lasting over a few days is seen, in general, but not always,  showing slow rise and fast decrease, with 2-day binned data-set. A slow rising trend in flares suggests fast cooling of electrons and a less stochastic acceleration process \citep{kapanadze2016}. A prolonged activity in X-ray band showing chaotic behavior  in the light-curve is noticed, when  the source was peaking at GeV, $\gamma-$ray energies. UV and optical peaks were seen to occur later as compared to X-rays. 

\subsubsection{Correlation in the flux variations}
 It is clear from Figures \ref{fig:fig1} \&  \ref{fig:fig2} that while flux starts rising first at lower energies (V, UV, X-ray) during outburst 2, it peaks first at  $\gamma$-rays followed by  X-ray, UV \& V bands. To check if the variations in these bands are correlated, we used discrete correlation function (DCF) which was first introduced by  \citep{Edelson1988} and generalized by \citep{Hufnagel1992} to include a better error estimate.  A brief description of the method is given by \citet{Tornikoski1994} and \citet{Hufnagel1992}. A variant of DCF is the zDCF \citep{Alexander2014} which corrects for various biases of the DCF method by employing equal population binning and Fishers z-transform. Lags have been computed based on maximum likelihood criterion satisfying one sigma confidence interval. 

Figure \ref{fig:fig3} shows three columns consisting of three panels each. Top two panels show light curves at two different energies while bottom panel shows discrete correlation between them.  In the correlation study on the present data sets, we notice that due to multiple overlapping flares in almost all the bands, and particularly erratic, elevated flux in X-rays, the correlations between various fluxes are not very strong. We could not get a clear correlation with X-ray vis-a-vis other bands. However, as is reflected by light-curves also, UV and optical fluxes are correlated (figure3, column 3), with UV leading optical by about a few hours. Similarly, $\gamma$-ray versus V and UV are correlated with $\gamma-$ray leading by 20 and 18 days, respectively. It can be clearly seen that events in $\gamma-$rays are fairly correlated with events in UV, optical and $\gamma-$ray variations. Therefore, the high energy $\gamma-$ray emission was followed by emissions at lower frequencies, X-ray, UV and optical, in general. However, there are instances when $\gamma-$ray emission lags behind low energy emissions, which can be explained based on light-travel time arguments and/or differently aligned emissions regions with respect to LOS.

 In case of HBLs, low energy emissions (IR to soft X-rays) are produced by the relativistic electrons as synchrotron radiation while the  high energy emission is expected to be generated through Inverse Compton (IC) process under which synchrotron photons are up-scattered by the same population (synchrotron) of electrons which produced them.  Generally, it is the most preferred scenario known as one-zone SSC that is capable of explaining SED of the high energy peaked blazars (HBL) and has been used in other studies \citep [and references there-in]{Bottacini2010}. 

 Since significant lags were noticed between high energy $\gamma-$rays and other low energy emissions, opacity effects of the  turbulent medium inside the jet could be responsible. In this case, the higher energy emission would occur first followed by low energy emission, as longer wavelengths are more susceptible to opacity effects compared to shorter ones.
The second explanation could be if the emissions are generated in different regions and/or are aligned differently to the line of sight of an observer. For example, if emission region emitting at higher energy is oriented closer to observer's line of sight as compared to other regions emitting at lower frequencies, then high energy emission would be more strongly Doppler boosted (due to relativistic effects) and will show faster variations in the light-curve \citep{MK97, Finke2008}. The low-energy emitting region being slightly away from observer's line of sight, would show delayed emission. This could be the reason that an activity is first seen at higher energies followed by lower energies and still appears to be correlated.

Nevertheless, the origin and prolonged activity in X-rays, as mentioned above, is still intriguing. The lack of connection between optical/UV and X-rays, very low degree of polarization (random), longer cooling timescales in X-ray light curves as noticed for Mrk 421 \citep{balokovic2016} suggest contributions from multiple emission regions. \citet{raiteri2015} suggested a complex UV and X-ray behavior for PG 1553+113 using multi-wavelength WEBT campaign data. A recent study by \citet{Cavaliere2017} suggested the presence of an extra keV synchrotron component during particle progressive acceleration along with canonical optical to GeV emissions. It would, therefore, be very interesting to look for the processes and regions leading to the origin of  X-rays and their relationship with other energy-bands during outburst and longer quiescent states to understand emission mechanisms at work inside the jet. 

\subsubsection{Synchrotron cooling time scale and magnetic field strength}

In the earlier section, we noticed that during second major outburst, the optical emission was delayed by a few hours with respect to the emission in UV band.  Now, since 1ES 1959+650 is an HBL, both the emissions are generated  by synchrotron process in which electrons are accelerated to the relativistic velocities, which then cool down, radiating at various frequencies- higher frequency emission being emitted first due to faster cooling rate.  Therefore,  the  time lag between emissions at two frequencies can be taken as the difference in the radiative cooling time scales of the population of electrons emitting at those frequencies \citep{UrryCM1997, baliyan1996}.
We can, therefore use the delay between the UV and optical emissions as cooling timescale of the synchrotron electrons and estimate the magnetic field. Using the expression for cooling time scale  \citep{UrryCM1997}, 

\begin{equation} \label{eq:2}
t_{lag} \approx t_{cool}  = 2.0 \times 10^{4} \sqrt{\frac {\delta}{(1+z)}}  B^{-3/2}  (\nu_{15}^{-1/2} )  sec
\end{equation}
where, B is magnetic field in Gauss, $\nu_{15}$ is the  frequency in $10^{15}$ Hz, $\delta$ $=$ 15 \citep{MK97} and $t_{lag}$ = 2.34 hrs. This gives us a magnetic field estimate of 4.21 G which is on the higher side in such systems.

\subsubsection{Estimation of the sizes of emission regions}
\label{sec:sec3.1.1}
The central regions of the AGN are very compact in size and can not be resolved by any existing facility. The variability property provides a tool to explore those deeper regions. We have noticed that there are several flares with short time scales. The shortest time scale of variation provides an upper limit to the size of emission region at that particular waveband, based on the arguments of light-travel time. 

In $\gamma-$rays, doubling time scale is the time period when flux doubles its initial value with more than 3$\sigma$ significance and we used it to estimate the upper bound of an emission region, $R_{\gamma}$ \citep{saito2015}. Based on the light-travel causality relation, we estimated R from,

\begin{equation}
\label{eq:3}
R \textless \frac{\delta c \tau_{d}}{(1+z)}
\end{equation}

where, R is the radius of the emission region, $\delta$ is taken as 40 \citep{aliu2014} for $\gamma-$rays, $\tau_{d}$ = 1 day is the flux doubling timescale and z= 0.048, redshift of the source. The $\gamma-$ray emission size is estimated as $R_{\gamma} \textless 9.89 \times 10^{16}$ cm. 
For the optical emission region,  we used  variability timescale of 4.5 hrs as a characteristic timescale and $\delta =15$, to estimate the upper limit to the size of the emission region as $R_{op}$  $\leq$ 3.71 x $10^{16}$ cm.  Thus $\gamma-$ray emission region is of the same order but about 3 times larger in size than optical region size.
The emission region sizes  obtained suggest that the $\gamma-ray$ and optical emission might be co-spatial in nature. However, one has to be cautious as due to the larger bin size in case of $\gamma-$rays  as compared to optical, it is difficult to get an accurate value for doubling time scale.

The location of the $\gamma-$ray production site is not very well known for all the sources, in general. But, for a few sources many authors reported location of  $\gamma-$ray emission region based on VLBI and high energy (GeV to TeV) simultaneous data-sets \citep{ Agudo2011, marscher2014}. Based on these observations, respective models predict that the $\gamma-$rays are produced close to the standing radio core (at 1-10pc) as $\gamma-$rays cannot escape from the vicinity of black hole due to photon absorption effects.

To determine the distance of the $\gamma-$ray emitting region from the central source, we need to know the opening angle close to the base of the jet,  Doppler factor ($\delta$), flux doubling time scale ($\tau_{d}$) and redshift (z).  The jet opening angle for blazars is generally less than 1 degree due to small viewing angle  \citep{Jorstad2005}, in general. Therefore, using the jet opening angle close to 1 degree, which is the upper limit of opening angle for BL Lacs \citep{Pushkarev_iX_2012}, we estimate the distance to the location of $\gamma-$ray emission from central engine, using following equation as,

\begin{equation}
d= \frac {\delta c \tau_{d}} {(1+z)\theta_j}
\end{equation}

The location of high energy $\gamma-$ray emission region from central SMBH is estimated as d = 1.72 pc. The result indicates that the location of  $\gamma-$ray production site is close to the standing shock (sub-mm core) inside the jet. Since the jet opening angle  for BL Lacs is difficult to measure due to several reasons, for eg., jet bending at parsec scale or faint emission where the jet bends etc. \citep{rector2003}, and the different values are reported by many authors \citep{ rector2003, Jorstad2005, Lister2011}, hence, the results should be considered with a caution. 

\subsection{Twin flares during October 2015 outburst: Optical/$\gamma$-ray light-curve}
\label{sec:sec3.1.2}

Figure \ref{fig:fig2} (left side;  top to bottom) shows $\gamma$-ray to radio band light-curve with optical polarization for 1ES 1959+650 during MJD 57250 - 57460 constructed using data from space-based instruments (LAT/XRT/UVOT), from MIRO and Steward Observatory (R band converted to V-band as given in \citet{Tagliaferri2003}). On a careful look, a twin peak structure is noticed in optical as well as in  $\gamma$-ray light-curve during outburst and post-outburst phase, respectively. Optical twin flares peak around 2015 November 18,  MJD 57344 (R = 14.45 $\pm$ 0.02 mag) and December 5, MJD 57361 (R = 14.52 $\pm$ 0.02 mag); whereas $\gamma$-ray twin flares were recorded around 2015 November 30, MJD 57360 and December 5, MJD 57365.5, both showing similar flux levels. It appears that the emission in $\gamma$-rays and optical are correlated with each other peaking almost at same time.

From the onset of outburst, 1ES 1959+650 brightened gradually during October 2015 with V = 14.85$\pm$ 0.02 mag on MJD 57306 (2015 October 11). On  October 13, it brightened by 0.08 mag within a day, decaying by 0.06 mag during MJD 57309 and MJD 57310. On the later date, it brightened by 0.07 mag in about four hours, to 14.76 $\pm$0.02 mag. After that, the source went into low-flux state and started dimming, reaching 14.81 (0.02) mag on MJD 57311 (October 16, 2015). However, on MJD 57344 (November 17, 2015) our observations detected the source in its brightest level during whole 2015 with 14.45 $\pm$0.02 mag.

The twin peaks in optical have been reported in a few HBLs \citep{sokolovMarscher2004}.  The flux enhancement in the sources like blazars is well explained by shock models, kink models etc. Here a  propagating shock in a random magnetic field plasma \citep{bland1979, marschergear1985} hits the Mach disk which leads to the formation of double structured feature in the light-curve. HBLs  1ES 0229+200, 1ES 0502+675, 1ES 2344+514 etc. are seen with such double peaks in optical, with or without periodicity present in their light curves \citep{kapanadze2010}. On the other hand, in the shock-in-jet model, at the shock front, two regions contribute to the emitted radiation - emission from forward and reverse shocks which could be responsible for short duration twin structures in the optical light curve. On the other hand, in sufficiently magnetized environments, kink instabilities can efficiently convert the magnetic energy into bulk kinetic and thermal energy in the jet . When the shock propagates down the jet through kinky nodes, it illuminates them along its path before getting dissipated as described by \citet{zhang2016} using relativistic magneto hydrodynamic simulations.

\subsection{DP and PA change during October 2015 outburst}
\begin{figure*}
\centering
\caption{(Left): Detailed light curves in $\gamma-$ray, X-ray, UV, optical and radio with optical polarization (DP) and position angle (PA) for the duration between  MJD 57250 - 57450 (August 2015 to March 2016).  (Right) : Zoomed version of the light curve in $\gamma-$ray and optical where rapid changes in DP and PA occurred within five days (MJD: 57306.5 - 57311.5).  \label{fig:fig2}}
\includegraphics[width=8.5cm,height=10cm]{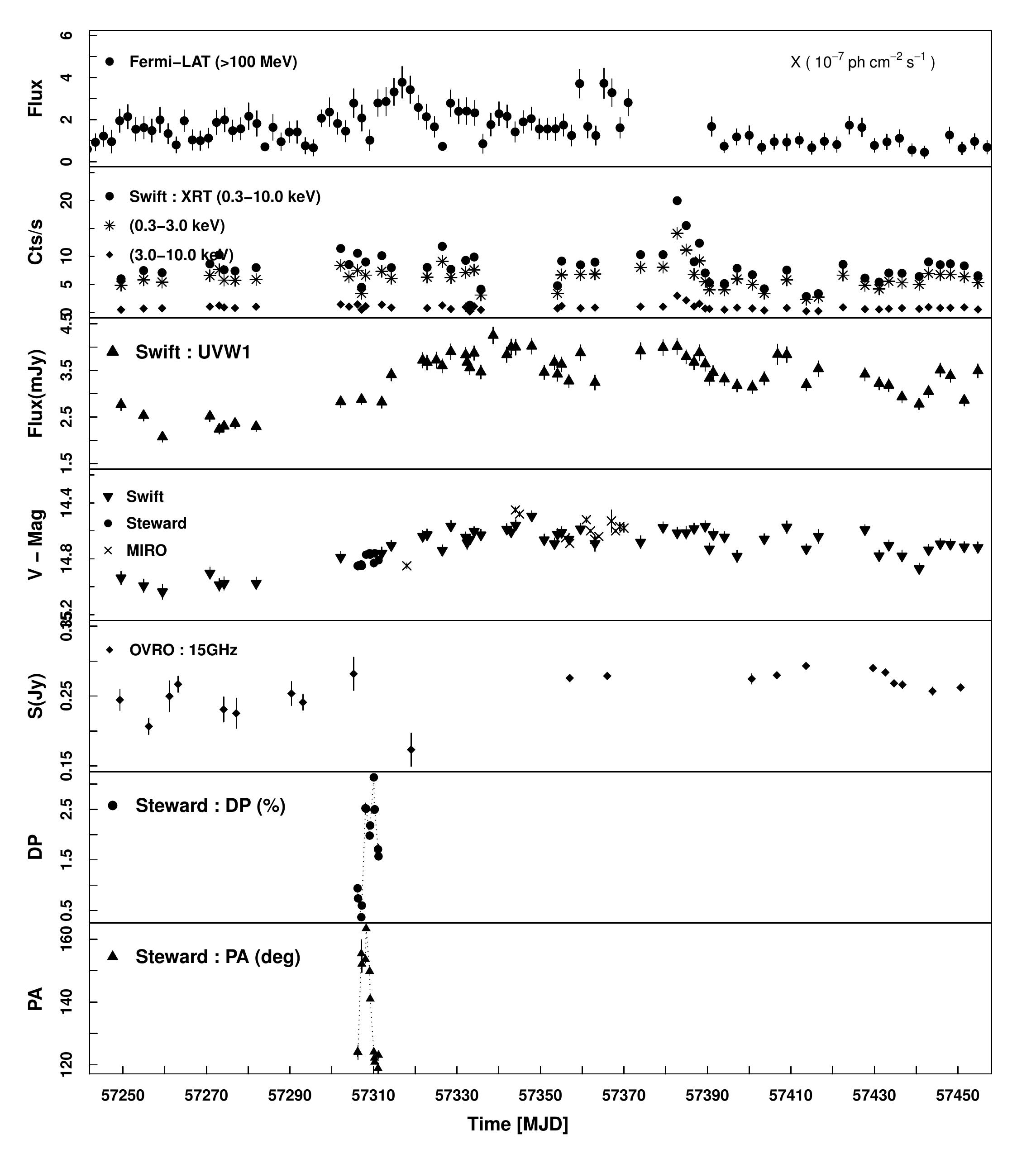}
\includegraphics[width=8.5cm,height=10cm]{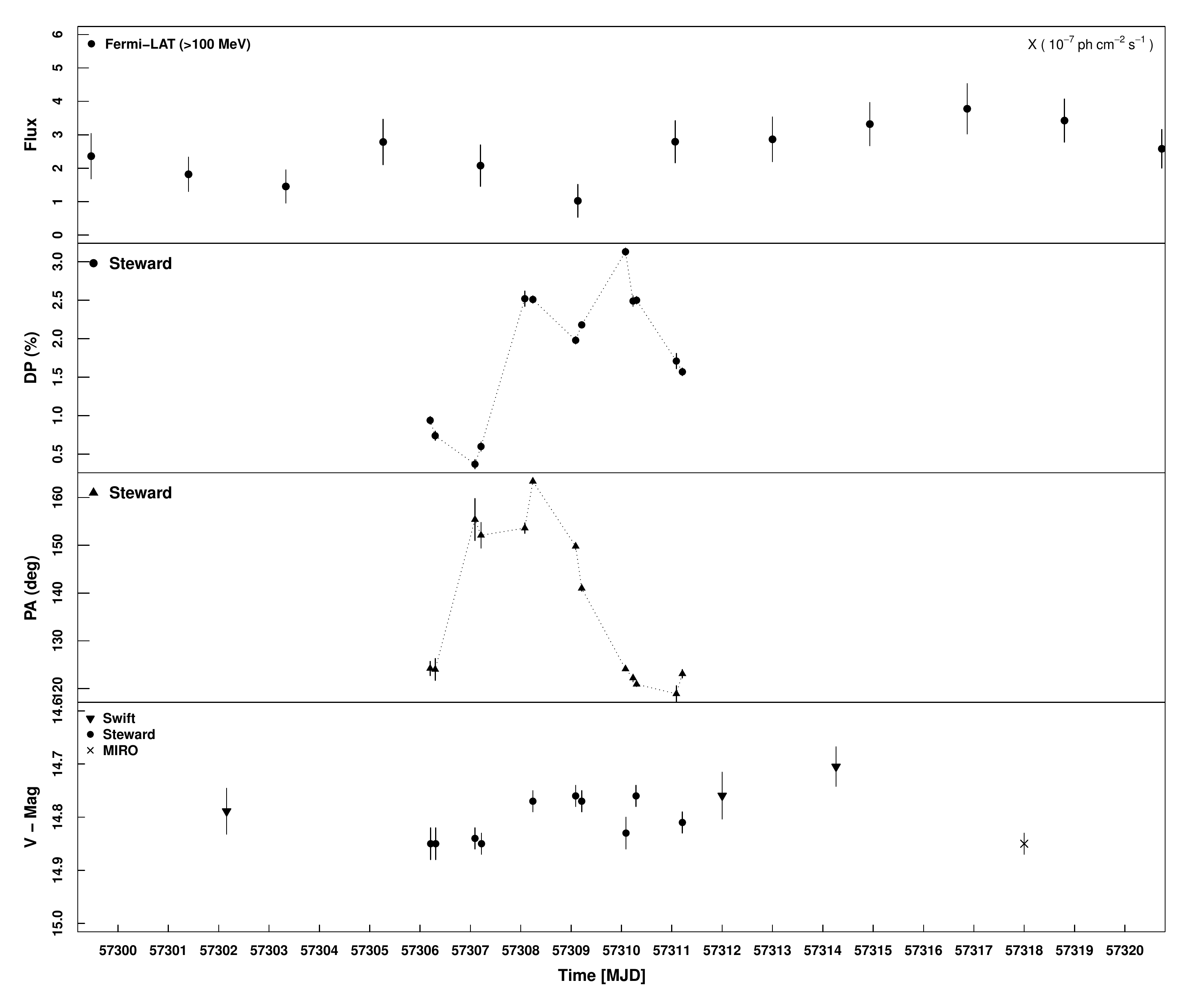}
\end{figure*}

\begin{figure*}
\centering
\caption{Discrete correlation plots of 1ES 1959+650, during October 2015 outburst, is shown between different energies. Top two panels shows light curves in different energies (X-axis : MJD; Y-axis : Flux/Mag) while the bottom panel shows correlation between them  (X-axis: Lag in days; Y-axis: zDCF). \label{fig:fig3}}
\includegraphics[width=5cm,height=10cm]{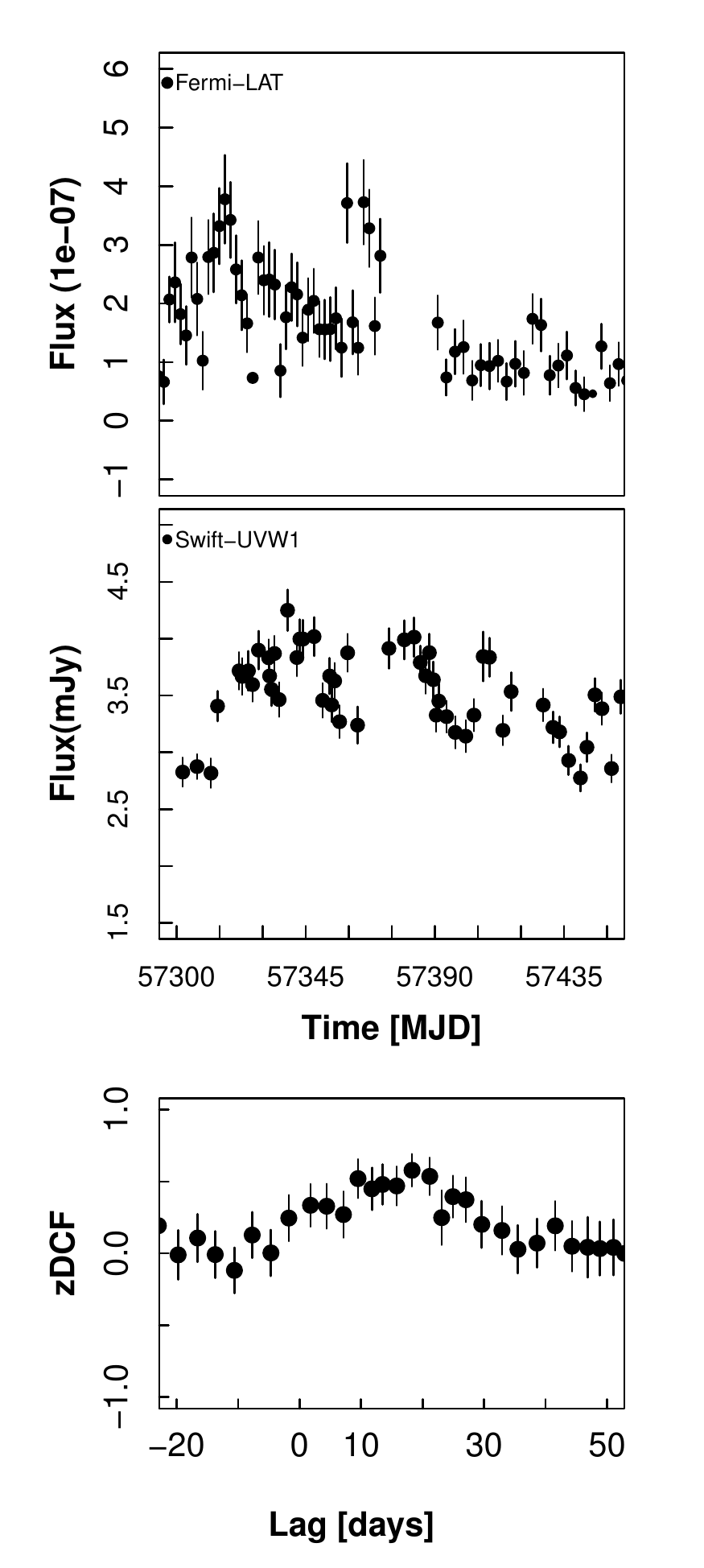}
\includegraphics[width=5cm,height=10cm]{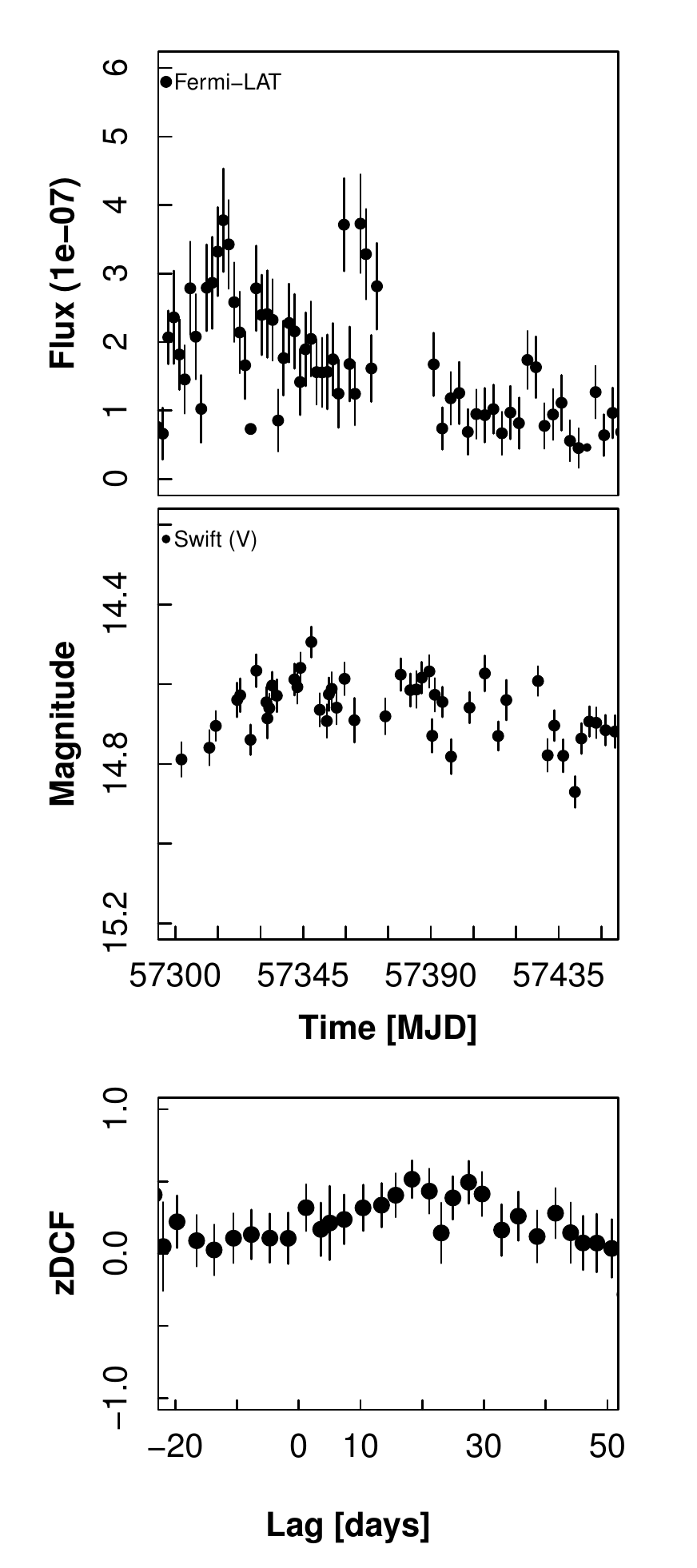}
\includegraphics[width=5cm,height=10cm]{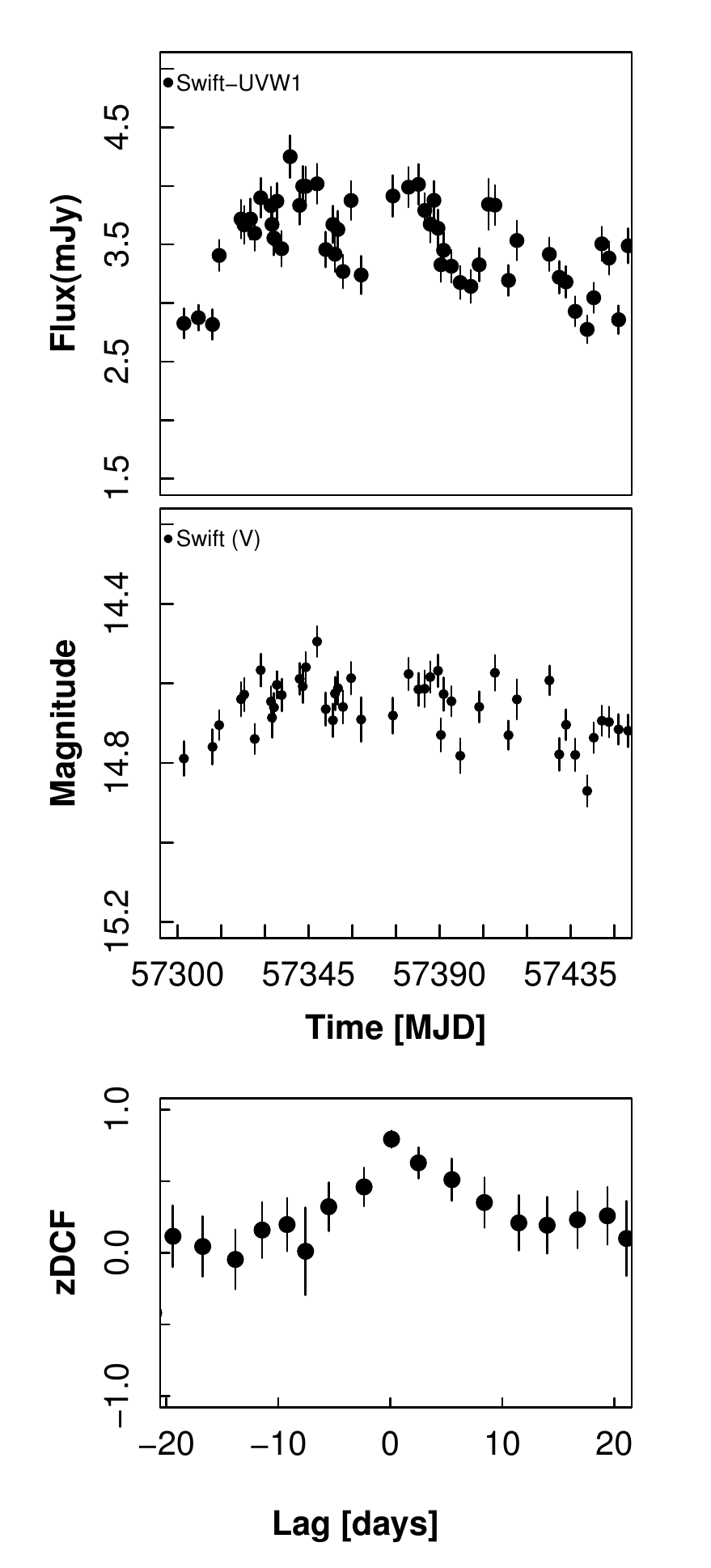}
\end{figure*}

It is interesting to see how the polarization behaved during multi-frequency outburst during  October 2015 in 1ES 1959+650. The Steward observatory optical polarization and position angle (DP and PA) data are plotted in the bottom two panels of Figure \ref{fig:fig2} and, with more clarity, in Fig.\ref{fig:fig2} (on the right) along with  R-band data for October 11 - 16, 2015 (MJD 57306 - 57312). During October 12 -13, 2015 DP along with R-band brightness increased sharply (DP:~0.3\% to 2.5\%, R mag: 14.37 to 14.31) followed by 10 degree change in PA to 153 degree . While brightness and DP remained at higher levels (DP: 3.13 - 1.57, R-mag: 14.3 -14.37), PA, after successive rotations by 20 degree on next two days, settled down around a value of 120 degree. DP is maximum when the source is  brightest during these observations. Variations in DP and R-band magnitudes are also seen at intra-night time scales as well. On October 15, 2015, 1ES 1959+650 brightened by 0.07 mag in R (about $3\sigma$) in about 4.8 hrs, while DP changed by 0.64\% $(> 8\sigma)$ within 3.6 hrs.
 
Here, we discuss multi-wavelength flare patterns with changes in optical polarization features. Figure \ref{fig:fig2} clearly shows that DP and PA significantly changed during October 2015 outburst when the source was in very high flux state across the whole electromagnetic spectrum (EMS). While the source was slowly brightening in optical, X-ray and $\gamma$-ray, the degree of polarization changed by 2.8 \% within 4 days i.e., from MJD 57307 (DP: 0.3 \%) to MJD 57310 (DP: 3.0 \%). This increase in DP is  followed by a change in the position angle of polarization by almost 80 degrees within six days duration. The 15 GHz radio flux reached its peak and then decreased sharply during these changes in optical polarization features, after which flux in $\gamma$-rays, followed by optical, UV and X-rays peaked. Perhaps, all this coincided with injection of fresh plasma in the jet which led to flaring in radio, gamma-rays and other bands. The situation well suits for the case of emission feature moving down the jet, interacting with the standing shock which results in compression of plasma, alignment of magnetic field resulting in polarization changes and acceleration of charged particles. These physical processes lead to enhanced emission at all the frequencies and increased degree of polarization \citep{marscher2014}. Unfortunately, the polarization data does not cover the domain of flares reaching their peaks.

\subsection{Injection of a new component in the jet}
In order to understand  the multi-frequency connection of the major outburst activity in 1ES 1959+650 during October 2015, 15 GHz OVRO data 
and  Astronomers Telegrams (ATels) on TeV activity reported during this outburst are considered. During MJD 57303-57304, the onset of TeV activity \citep{mukherjee2015_atel_8148}  showing much harder spectra, $\alpha_{(0.2-7) TeV}$ = 2.5, was followed by a significant enhancement in 15 GHz radio flux, $F_{radio}$ =  0.28 $\pm$ 0.02 Jy) and $\gamma$-ray $(> 100 MeV ; F_{\gamma} = (2.27 \pm 0.68) \times$ 10$^{-07}$ ph cm$^{-2}$ s$^{-1})$ . After six days of flux rise in radio and gamma-rays, the source showed historically highest X-ray fluxes on MJD 57311.99 i.e., (10.13 $\pm$ 0.10) cts/s in (0.3-10.0) keV and (7.34 $\pm$ 0.09) cts/s in (0.3-3.0) keV. A quasi-simultaneous enhancement in GeV, $F_{\gamma} = (2.86 \pm 0.67)$ $\times$ 10$^{-07}$ ph cm$^{-2}$ s$^{-1}$ on MJD 57313;  $F_{\gamma} = (3.77 \pm 0.75)$ $\times$ 10$^{-07}$ ph cm$^{-2}$ s$^{-1}$ on MJD 57316.86 and (7.99 $\pm$ 0.06) cts/s in X-ray (0.3-10.0) keV on MJD 57314.28 were reported.  
The source was detected with  the brightest ever in X-ray flux on MJD 57326.48 with exceptionally high count rates i.e., (11.81 $\pm$ 0.09) cts/s in (0.3-10.0) keV and (9.12 $\pm$ 0.08) cts/s in (0.3-3.0) keV. The source was dimming in GeV ($F_{\gamma} = (2.05 \pm 0.54$) $\times$ 10$^{-07}$ ph cm$^{-2}$ s$^{-1}$ on MJD 57347.80) but brightening up in optical, showing V= 14.94 $\pm$ 0.03 mag on MJD 57347.91. We also noticed flux enhancement in radio by almost 0.1 Jy, showing $F_{radio}$ = (0.275 $\pm$ 0.003 Jy)  on MJD 57357, followed by GeV activity on MJD 57365.20 ($F_{\gamma} = (3.73 \pm 0.72$) $\times$ 10$^{-07}$ ph cm$^{-2}$ s$^{-1}$). 

The almost continuous enhancement in flux during October 2015 outburst, at different frequencies, suggests a scenario of continuous injection of fresh relativistic particles near the origin of the jet, that accelerates the particles (electrons) up to relativistic energies, which, in turn cool down by emitting  higher energy photons followed by the emissions at lower frequencies. This can also explain a mild bluer when brighter color seen in the present study. However, color remains same with time, within errors, except during flaring when it shows a BWB behavior, typical of HBLs. A quasi-simultaneous enhancement in the high energy $\gamma$-ray and radio flux, perhaps, may be associated with the  presence of a standing shock feature or core. The emission in $\gamma-$rays is supposed to arise from the seed photons in the acceleration and collimation zone inside the jet, where all particles (here, electrons) spiral in helical magnetic field and emit in VHE gamma rays \citep{marscher2008}. The radio emission is opaque upstream the stationary core due to synchrotron self absorption effects, and becomes transparent only when the emission feature interacts with the radio-core.  

During October 2015 outburst, correlated flux in radio and TeV energies is noticed using publicly available data and a few published reports. A quasi-simultaneous flux variability is also noticed by \citet{tagliaferri2008} and \citet{hayashida2008} along-with many other authors who  reported a correlated activity between highly variable strong $\gamma-$rays and radio frequencies \citep{ramakrishnan2015} with an evidence of the  ejection of a super-luminal knot from radio core using VLBI analysis \citep{jorstad2001,schinzel2012,jorstad2015}. Their study suggests that if the origin of seed photons responsible for  enhanced gamma-rays is from the radio core (mm-core), then it would be due to the interaction of the moving shock with the standing core.

\section{Summary and conclusions}
We presented the analysis of the multi-wavelength data from Fermi and Swift, available publicly, MIRO and Steward Observatory optical data  for HBL 1ES 1959+650 during the year 2015-2016, covering two major outbursts. We also made use of available optical polarization data to discuss October 2015 outburst in detail. The source was active in all the energy bands and showed significant flux enhancements during most of the period covered in this study. It is worth noting that 1ES 1959+650 exhibited highest ever flux in $\gamma-$rays and more than 20 counts per second in X-ray emission, making it third source after Mrk421, Mrk501 having such high counts. MIRO data showed the source in its brightest state (V = 14.45 $\pm$ 0.03) during 2015.

A mild indication of the optical intra-night variability is seen on one of the nights during October 2015 when the source brightens by 0.07 mag in 4.5 hrs. Also, the source exhibited short-term variability (over a few days) with a significant ($>$ 0.3 mag in V band) variability amplitude. 
The first outburst (March 2015) was characterized by the emission in $\gamma-$rays, perhaps, to follow those at lower frequencies with origin from synchrotron radiation, which is explainable via SSC mechanism. The second outburst (October 2015) was rather complex with the 15GHz radio emission peaking first, just around the time optical polarization features changed rapidly. It was followed by the peaks in $\gamma-$ray flux, X-ray, UV and optical emissions. UV and optical variations were delayed by about 20 days with respect to those in $\gamma-$rays. It appears that various emission regions were aligned differently resulting in varying Doppler boosting of the flux and variability time scales. However, since $\gamma-$ray flux peaked just after radio, followed by emissions at lower frequencies, the processes could be related to the  injection of fresh plasma in the jet. We estimated  magnetic field strength (B $=$ 4.21 G) by using time lag between UV and optical emission as synchrotron cooling timescale. The emission region sizes for $\gamma-$ray and optical were estimated using shortest time scales of variability which were found to be  of the order of $\approx \ 10^{16}$ cm. The $\gamma-$ray emission region appears to be located at a distance of 1.72 pc from central SMBH, which is close to the standing shock feature in the jet. As per the Astronomers' Telegrams (quasi-) simultaneous enhancement in radio and TeV flux were noticed during  MJD 57303-57304, indicating to the injection of  a new component in the jet which propagates down the jet interacting with standing conical shock (or radio core). It leads to quasi-simultaneous emissions at almost all the frequencies. The long term multi-frequency study suggests a mild bluer-when-brighter trend in flux, which is relatively stronger during flaring.  

\section*{Acknowledgements}

This work is supported by the  Department of Space, Government of India. Data from the Steward Observatory spectro-polarimetric monitoring project were used. This program is supported by Fermi Guest Investigator grants NNX08AW56G, NNX09AU10G, NNX12AO93G, and NNX15AU81G. This research made use of Enrico, a community-developed Python package to simplify Fermi-LAT analysis \citep{sanchezdiel2013}. This research has made use of data from the OVRO 40-m monitoring program \citep{richard2011} which is supported in part by NASA grants NNX08AW31G, NNX11A043G, and NNX14AQ89G and NSF grants AST-0808050 and AST-1109911.

\newpage

\newpage

\begin{deluxetable*}{ccccccccccc}
\tablecaption{The optical data on 1ES 1959+650 obtained from MIRO using 1.2m and 0.5 m telescopes and from Steward Observatory archive for the duration October $-$ December 2015. Column 1 and 2 are the epochs of the observations (date  and MJD format), Column 3, 4 and 5, 6 are optical R-band and V-band magnitudes with associated errors, column 7,8 and 9,10 give degree of polarization and position angle with respective errors, as obtained from the telescopes listed in column 11.\label{tab:table1}}

\tablehead{Date & MJD &R &$\sigma_R$&V & $\sigma_V$& DP($\%$) & $\sigma_{DP}$ & PA (deg) & $\sigma_{PA}$&Observatory}

\startdata
11-Oct-2015	&57306.30 			&--		&--		&--		&--		&0.74 	&0.06 	&124.0 	&2.3		&Steward 	\\	
			&57306.70			&14.39	&0.03	&--		&--		&--		&--		&--		&-- 		&Steward \\
			&57306.81			&14.39	&0.03	&--		&--		&--		&--		&--		&-- 		&Steward \\
12-Oct-2015	&57307.09 	 		&--		&--		&--		&--		&0.37 	&0.06 	&155.4 	&4.4 		&Steward \\
			&57307.21 			&--		&--		&--		&--		&0.60 	&0.06 	&152.1 	&2.7 		&Steward \\		
			&57307.59			&14.38	&0.02	&--		&--		&--		&--		&--		&--		&Steward\\
			&57307.71			&14.39	&0.02	&--		&--		&--		&--		&--		&--		&Steward \\
13-Oct-2015	&57308.08			&--		&--		&--		&--		&2.52 	&0.10 	&153.6 	&1.1		&Steward  \\
 			&57308.24 			&--		&--		&--		&--		&2.51 	&0.05 	&163.4 	&0.5 		&Steward \\
			&57308.74			&14.31	&0.02	&--		&--		&--		&--		&--		&--		&Steward \\
14-Oct-2015	&57309.09 	 		&--		&--		&--		&--		&1.98 	&0.05 	&149.8 	&0.7 		&Steward\\
			&57309.21  			&--		&--		&--		&--		&2.18 	&0.04 	&141.0 	&0.6 		&Steward\\
			&57309.59			&14.30	&0.02	&--		&--		&--		&--		&--		&--		&Steward \\
			&57309.71			&14.31	&0.02	&--		&--		&--		&--		&--		&--		&Steward \\
			
15-Oct-2015	&57310.08 			&--		&--		&--		&--		&3.13 	&0.05 	&124.1 	&0.4		&Steward \\
 			&57310.23 	 		&--		&--		&--		&--		&2.49 	&0.07 	&122.2 	&0.8 		&Steward 	\\
			&57310.59			&14.37	&0.03	&--		&--		&--		&--		&--		&--		&Steward \\
			&57310.78			&14.30	&0.02	&--		&--		&--		&--		&--		&--		&Steward \\
16-Oct-2015	&57311.09 			&--		&--		&--		&--		&1.71 	&0.10 	&118.9 	&1.7		&Steward 	\\
			&57311.21 			&--		&--		&--		&--		&1.57 	&0.05 	&123.1 	&0.9 		&Steward\\
			&57311.71			&14.35	&0.02	&--		&--		&--		&--		&--		&--		&Steward \\
23-Oct-2015	&57318				&14.39	&0.02 	&14.80	&0.03	&--		&--		&--		&--		&MIRO (1.2m)\\
17-Nov-2015	&57344				&13.99	&0.02	&14.48	&0.02 	&--		&--		&--		&--		&MIRO (1.2m)\\
18-Nov-2015	&57345				&14.02	&0.02	&14.45 	&0.01 	&--		&--		&--		&--		&MIRO (1.2m)\\
29-Nov-2015	&57356				&14.19	&0.01	&--		&--		&--		&--		&--		&--		&MIRO (1.2m)\\
30-Nov-2015	&57357				&14.23	&0.01	&--		&--		&--		&--		&--		&--		&MIRO (1.2m)\\
04-Dec-2015 	&57361 				&14.06	&0.02  	& 14.56	&0.01	&--		&--		&--		&--		&MIRO (0.5m)\\
05-Dec-2015 	&57362	 			&14.14	&0.01  	& 14.57	&0.03	&--		&--		&--		&--		&MIRO (0.5m)\\
06-Dec-2015 	&57363 				&14.20	&0.02  	& 14.61	&0.05	&--		&--		&--		&--		&MIRO (0.5m)\\
07-Dec-2015 	&57364 				&14.18	&0.02  	& 14.58	&0.04	&--		&--		&--		&--		&MIRO (0.5m)\\
10-Dec-2015 	&57367 				&14.07	&0.08  	&   --      	&--		&--		&--		&--		&--		&MIRO (1.2m)\\
11-Dec-2015	&57368 				&14.14	&0.01	&   --      	&--		&--		&--		&--		&--		&MIRO (1.2m)\\
12-Dec-2015	&57369				&14.11	&0.03	&   --      	&--		&--		&--		&--		&--		&MIRO (1.2m)\\
13-Dec-2015 	&57370 				&14.12	&0.03 	&   --      	&--		&--		&--		&--		&--		&MIRO (1.2m) \\
\enddata 

\end{deluxetable*}

\bibliographystyle{apj}
\bibliography{reference}

\clearpage
\end{document}